\documentclass[5p, twocolumn]{elsarticle}
\usepackage{geometry}
\usepackage{amsmath}
\usepackage{amstext}
\usepackage{amssymb}
\usepackage{hyperref}
\hypersetup{colorlinks=true, citecolor=red, urlcolor=red, linkcolor=red}
\usepackage{cleveref}
\biboptions{sort&compress}
\usepackage[table]{xcolor}
\definecolor{table-gray}{gray}{0.85}
\begin{document}
	
	\begin{frontmatter}
		\title{\textbf{Data center energy efficiency enhancement using a two-phase heat sink with ultra-high heat transfer coefficient}}
		\author[1]{Suhas Rao Tamvada}	
		\author[1]{Saeed Moghaddam\corref{cor1}} \ead{saeedmog@ufl.edu}
		\cortext[cor1]{Corresponding author}
		\address[1]{Mechanical and Aerospace Engineering, University of Florida, Gainesville, FL 32611, United States}
		\begin{keyword} data center cooling \sep heat transfer \sep phase change \sep boiling \sep two-phase flow
		\end{keyword}
		\begin{abstract}
This paper presents the latest progress on characterization of our membrane assisted phase-change heat sink (MHS) at conditions suitable for implementation in data centers (DCs). Experiments are conducted using water as the working fluid at a vapor space pressure ($P_{vapor}$) of 16 kPa, corresponding to a saturation temperature of $\sim$ 55$^{\circ}$C. This temperature is sufficiently lower than the silicon junction temperature of ~80$^{\circ}$C. As anticipated, the overall performance of MHS at sub-atmospheric pressure is lower compared to analogous tests at atmospheric pressure. In agreement with previous studies on MHS, the critical heat flux limit (CHF) increases with enhancement of the heat transfer area ratio ($A_r$) and liquid space pressure ($P_{pool}$). We report a maximum CHF of 670 W/cm$^2$ on a surface with an enhanced area ratio of 3.45, multiple times greater than the CHF reported hitherto by a comparable two-phase heat sink in literature. Heat transfer coefficients (HTC) as high as $\sim$ 1 MW/m$^2$-K are obtained. These record performance data along with unique characteristics of the MHS promise to greatly benefit next generation highly energy efficient DCs.
		\end{abstract}
	\end{frontmatter}

\section{Introduction \label{sec:intro}}
Proliferation of high-power density electronics has engendered the development of high-performance computing (HPC) to supplement the growing needs of artificial intelligence, big data, and cloud computing \cite{Li2020}. The energy demands of data centers (DCs) which house the IT hardware needed for these computing resources have been on the rise in the past decade and are projected to increase rapidly \cite{Kheirabadi2016, Alkharabsheh2015}. A myriad of factors influences DCs energy efficiency, including the DC size, compute density, ambient conditions, etc. \cite{Li2020}. While hyperscale DCs promise increased efficiency, a demand for low latency is promoting less efficient edge DCs (relatively smaller high-density facilities located close to the end-user). Energy use for thermal management of servers is a major component of the overall energy demand in a DC. A widely used metric for characterization of DCs energy efficiency is the power usage effectiveness (PUE) \cite{Garimella2013}. PUE is the ratio of the total power used in the DC to the power used by the IT equipment, given in \cref{eq:eq1}
\begin{equation} \label{eq:eq1}
PUE = \dfrac{P_{tot}}{P_{IT}}
\end{equation}
where $P_{tot}=P_{IT}+P_{T}+P_{e}$ is the total power supply to the DC, and $P_{IT}, P_{T},$ and $P_e$ denote power consumption by the IT equipment, thermal management systems, and the electrical power distribution system, respectively. Currently, the average PUE of a DC is about 1.7 \cite{Alkharabsheh2015} meaning that 70\% of the power flowing into DCs is utilized for auxiliary purposes, most important of which is the cooling system energy use \cite{Dayarathna2016}.

\begin{figure}[htp!]
	\centering
	\includegraphics[width=0.48\textwidth]{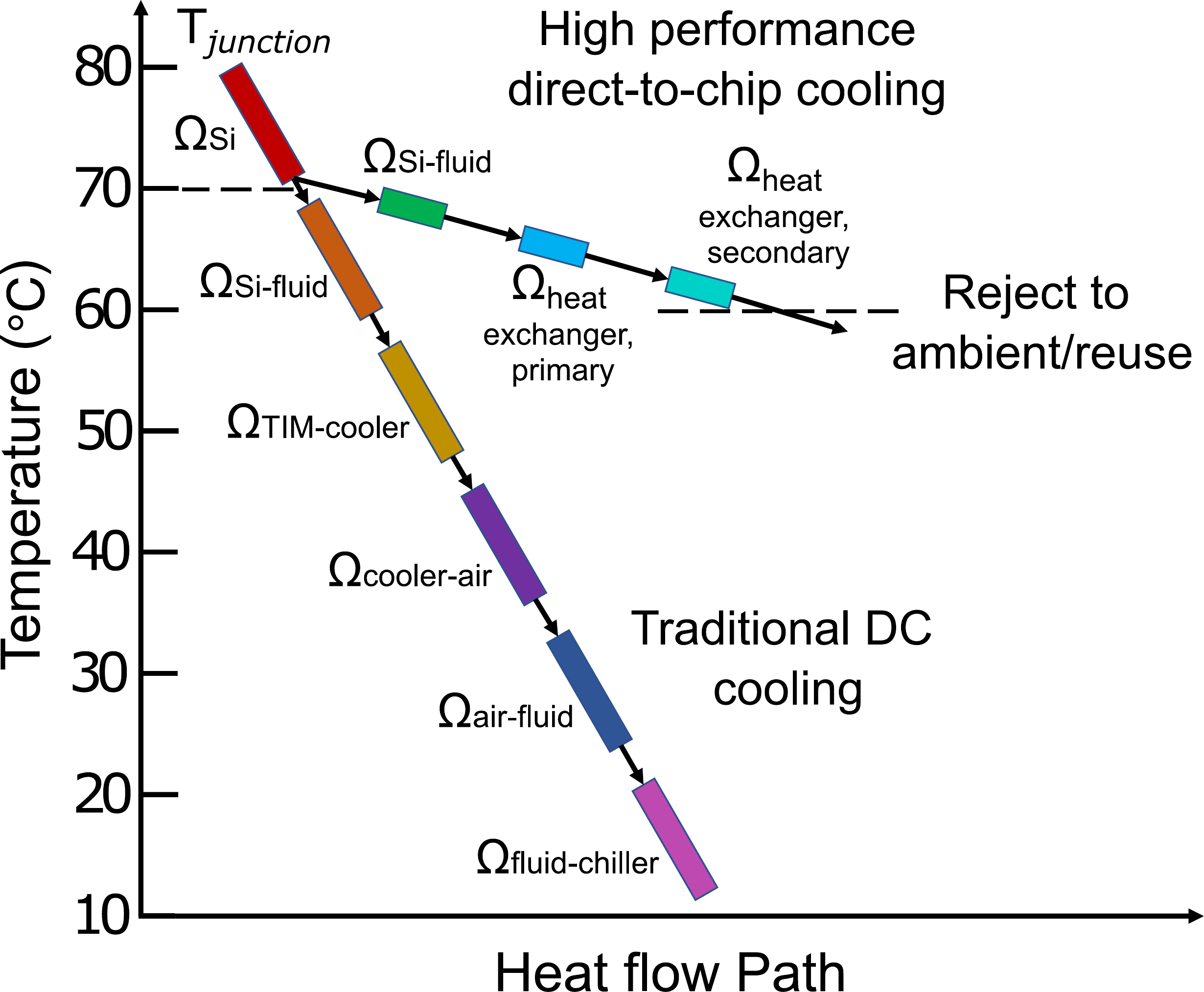}
	\caption{Thermal resistances encountered in a typical heat flow path in traditional data center (DC) cooling and a proposed heat flow path utilizing a high-performance heat sink at the chip level. The resistances between the thermal interface material, cooler, and fluids require the use of a chiller below ambient temperatures. Conversely, a high-performance heat sink mounted on the chip will eliminate multiple thermal resistances and need for power-hungry chillers. \label{fig:fig1}}
\end{figure}

The current technology involves cooling the server racks using raised floors and computer room air conditioning (CRAC) and air handling (CRAH) units \cite{Chu2019}. In a traditional air-cooled data center, thermal resistances ($\Omega$) from multiple components including the thermal interface materials (TIM), cooler, etc., between the junction and ambient (cf. \cref{fig:fig1}) make the heat flow path complex and require the use of chillers (7-15 $^\circ$C) to maintain an acceptable junction temperature. Alternatively, systems involving a high-performance heat sink at the chip level offers significantly lower thermal resistance and enable rejection of heat to the ambient at a higher temperature \cite{Marcinichen2014, Kandasamy2022}. In specific cases, this waste heat could be utilized in buildings and industrial units.

To realize such systems, there is an urgent need for development of robust high performance heat sinks. Thus far, single phase heat sinks have advanced electronic cooling limits in comparison to air cooling; however, the rapid increase in power density and miniaturization of electronics demand more efficient cooling. Utilizing phase change heat sinks is suggested to decrease DCs total cost of ownership (TCO) through reduction in operational expenditure (OpEx) \cite{Cui2017,Rasmussen2011}. Phase change heat sinks have demonstrated higher heat transfer coefficients (HTC) as a result of the phase change process and remove higher heat per working fluid mass flow rate relative to single phase cooling \cite{Matin2020, Moghaddam2006, Moghaddam2009}. Furthermore, two-phase heat sinks can enhance temperature uniformity across the chip \cite{Matin2021, Verma2021}. Nonetheless, most studies in literature on two phase heat sinks use water as the working liquid at atmospheric conditions where the saturation temperature of water is 100 $^\circ$C. These operating conditions are not representative of the conditions in a data center wherein the silicon junction temperature must be kept below $\sim$80 ℃ \cite{Xi2004}. Two phase heat sinks utilizing water must operate at saturation temperatures below the junction temperature. Operation at reduced saturation temperature results in decrease in vapor density and a drastic reduction in boiling performance \cite{ Van1975, Zajaczkowski2016, Michaie2017}, negating advantages of conventional two-phase over single-phase heat sinks \cite{McGillis1991}. To replace single phase heat sinks with better alternatives in applications requiring removal of high heat fluxes, the key question to be answered is – Can two phase heat sinks achieve better performance at lower pressures compared to single phase heat sinks? A major limitation of phase change cooling is the occurrence of the boiling crisis at high heat fluxes due to surface dryout and inadequate rewetting. The maximum heat flux that can be removed from a surface prior to dryout is known as the critical heat flux (CHF). Following the seminal study by Nukiyama \cite{Nukiyama1934}, on identifying CHF, many efforts have been directed towards understanding the limiting factors of CHF and increasing its value. Following an understanding that the CHF limit can be increased through efficient liquid delivery to the heater surface \cite{Bonilla1941, Cichelli1945, Kutateladze1948, Rohsenow1956, Zuber1959, Moissis1963, Bui1985, Kandlikar2001, Attinger2014}, efforts have focused on improving surface wettability \cite{Bui1985,Jo2011} and wickability (using micro/nanostructures) \cite{Attinger2014, Rahman2014, Chu2011, Chu2013}. Separation of liquid-vapor pathways through surface modifications have also resulted in enhancement of the CHF limit \cite{Kandlikar2017}.

In recent inventions \cite{Moghaddam2019, Moghaddam2021}, we have demonstrated that liquid and vapor pathways can be quite effectively separated through direct extraction of bubbles from above the boiling surface through a selectively permeable membrane. The membrane must be hydrophobic to hold liquid water adjacent to the boiling surface. We have coined the term Membrane assisted phase-change Heat Sinks (MHS) for this technology. Earlier demonstration of the device performance has shown unprecedented heat flux dissipations of 1-2 kW/cm$^2$, depending on MHS design and operating conditions, with water at atmospheric pressure \cite{Fazeli2017, Alipanah2020}, offering a promising alternative to conventional heat sinks at conditions relevant to DCs. In this study, we evaluate boiling performance of MHS at a sub-ambient pressure of 16 kPa. In the following sections, first, the operating principle of the MHS and the experimental setup are discussed. Next, results pertaining to CHF, HTC, surface temperature stability, and pressure drop are discussed and conclusions are drawn. Finally, the MHS performance is compared with other technologies and its thermal, device, and system level benefits are discussed. 

\section{MHS operating principle \label{sec:sec2}}
\begin{figure*}[ht!]
	\centering
	\includegraphics[width=\textwidth]{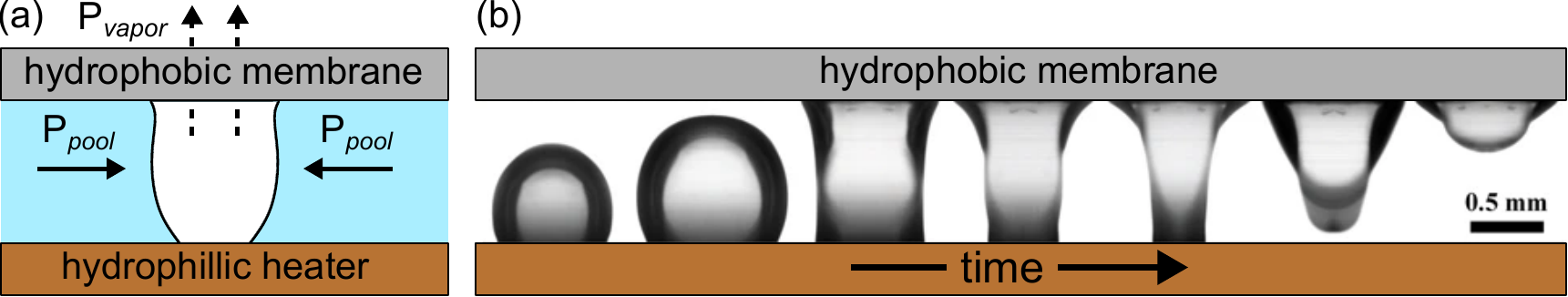}
	\caption{(a) Schematic of cross-section of a membrane-assisted heat sink (not to scale) shows the vapor expulsion mechanism through an omnidirectional pressure potential and vapor permeable membrane. (b) Visualization of a vapor bubble extraction process from a pool of water constrained by a hydrophobic membrane in an adiabatic test. \label{fig:fig2}}
\end{figure*}

The schematic in \cref{fig:fig2} (a) shows a cross-sectional view of the membrane-assisted heat sink discussed in this study. Unlike conventional heat sinks which have a liquid inlet and a two-phase flow outlet, a MHS contains only a single liquid inlet through which the working liquid is supplied to the heater surface at a pressure $P_{pool}$. A hydrophobic vapor permeable membrane placed $\sim$1 mm above the heater surface allows vapor exit from the liquid pool. This unique design confines the boiling liquid to the heater surface and induces an omnidirectional pressure on bubbles. With sufficient growth of a bubble, a vapor bridge is established between the heater surface and the membrane, leading to the liquid contact line on the membrane to recede (due to the membrane hydrophobicity) pulling and expelling the bubble from the heater surface.

\Cref{fig:fig2}(b) illustrates this phenomenon using an adiabatic test which consisted of injecting air into a pressurized pool of liquid at a constant flow rate of 100 ml/hr to replicate bubble growth. As the bubble departs, the surrounding liquid is delivered to the dry patch on the heater surface. This two-phase flow arrangement overcomes a key shortcoming of the pool boiling process – the evaporation/recoil pressure at the heater surface pushing the liquid away from the bubble nucleation site. This method of manipulating vapor discharge is shown to enhance multiple performance characteristics during boiling. For example, in addition to dissipating heat fluxes greater than one order of magnitude \cite{Fazeli2017, Alipanah2020} above the Kutateladze \cite{Kutaleladse1951} and Zuber \cite{Zuber1959} limit (119 W/cm$^2$ for a copper surface), rapid expulsion of vapor from the liquid pool greatly lowers two-phase pressure drop relative to conventional heat sinks \cite{Tamvada2021}. 
\section{Experimental studies}
\subsection{Heat sink test apparatus \label{sec:sec3-1}}
\begin{figure}[ht!]
	\centering
	\includegraphics[width=0.5\textwidth]{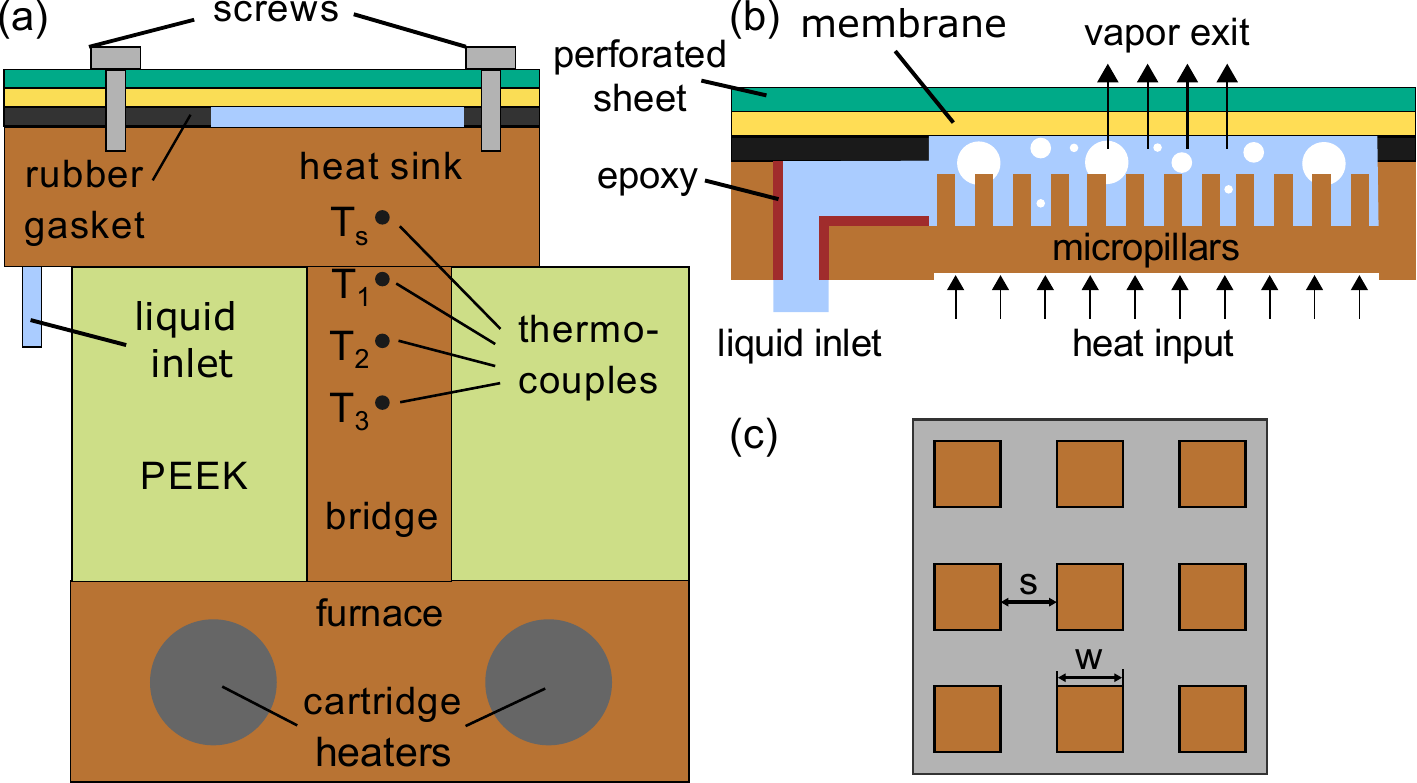}
	\caption{(a) Cross sectional view of the test device consisting of a furnace, bridge section insulated by PEEK, and a membrane-assisted heat sink (MHS). (b) cross sectional view of the MHS showing copper surface with micropillars, silicone spacer, hydrophobic membrane, and perforated metal support sheet. (c) surface micropillars machined with width \textit{w}, spacing \textit{s}, and height \textit{h} (out of plane, extruding out of paper).\label{fig:fig3}}
\end{figure}

The heat sink device consists of a 7$\times$7 mm$^2$ heater area with a liquid inlet channel of 1$\times$1 mm$^2$ cross sectional area, fabricated from superconductive copper 101 alloy (see \cref{fig:fig3}).

To ensure that boiling occurs only on the test surface, the side walls of the liquid delivery channel were covered with a 200-$\mu$m-thick layer of non-conductive epoxy by first machining a trench of area 1.2$\times$1.2 mm$^2$, filling the trench with non-conductive epoxy and re-machining the channel to a cross sectional area of 1$\times$1 mm$^2$ (see \cref{fig:fig3}(b)). The surface of the heater comprises of an array of micropillars machined on a micro-CNC as shown in \cref{fig:fig3}(c). Micropillars increase the total heater area in contact with the liquid, and the enhanced area ratio ($A_r$) is given by $A_r= 1+4wh\/(s+w)^2$ where \textit{w} is the width of the pillars, \textit{h} is the height of the pillars, and \textit{s} is the spacing between the pillars. By altering the width, height, and spacing between the pillars, the total heat transfer area and wickability of the surface can be enhanced by various degrees. Surface wickability, the ability of a structured surface to transport liquid through capillarity is known to enhance CHF \cite{Rahman2014}. Fazeli \& Moghaddam \cite{Fazeli2017} compared different methods of calculating surface wickability and found that in most cases, \textit{ad hoc} parameters are required to determine wickability. To avoid empirical assumptions, they defined wickability as the product of surface permeability ($\kappa_{wick}$) and capillarity ($P_c$) which we utilize in the current study. Micropillar dimensions, surface wickability, and corresponding area ratios are provided in \cref{tab:tab1}.

\begin{table}[htp!]
\begin{center}
\caption{Geometry, wickability, and enhanced area ratio $A_r$ of micropillars \label{tab:tab1}}
\begin{tabular}{c c c c c c}
\hline
Device & \textit{s}  & \textit{w}  & \textit{h} & $\kappa_{wick}P_c \times10^6$ & $A_r$ \\
& ($\mu$m) & ($\mu$m) & ($\mu$m) & (Pa/m$^2$) & \\
\hline
1 & N/A & N/A & N/A & N/A & 1  \\
\rowcolor{table-gray}
2 & 200 & 100 & 150 & 0.478 & 1.67 \\
3 & 300 & 350 & 350 & 0.754 & 2.16 \\
\rowcolor{table-gray}
4 & 200 & 150 & 500 & 0.757 & 3.45 \\
\hline
\end{tabular}
\end{center}
\end{table}

An 800 nm-thick oxide layer was thermally grown on the heater surface using a process used by Zhou and Yang \cite{Zhou2003} to ensure a consistent contact angle of $\sim 5^\circ$ in all experiments. The copper heat sink was brazed to a copper bridge section welded to a heating block using a high-temperature SnPb solder with a thermal conductivity of 57 W/m-K (see \cref{fig:fig3}(a)). The bridge section was insulated with Polyether Ether Ketone (PEEK) to minimize thermal losses. A 500-$\mu$m-thick rubber gasket was placed between the membrane and heat sink and fastened with the support of a perforated metal sheet. Three T-type thermocouples were installed equidistantly at 4 mm spacing within the copper bridge to determine the heat flux using a three-point backward difference Taylor’s approximation as given in \cref{eqn:eq2}.

\begin{equation} \label{eqn:eq2}
q'' = k \times \dfrac{3T_3 - 4T_2 + T_1}{2 \Delta x}
\end{equation}

A fourth thermocouple was placed below the surface to record the surface temperature.
\begin{figure*}[ht!]
	\centering
	\includegraphics[width=\textwidth]{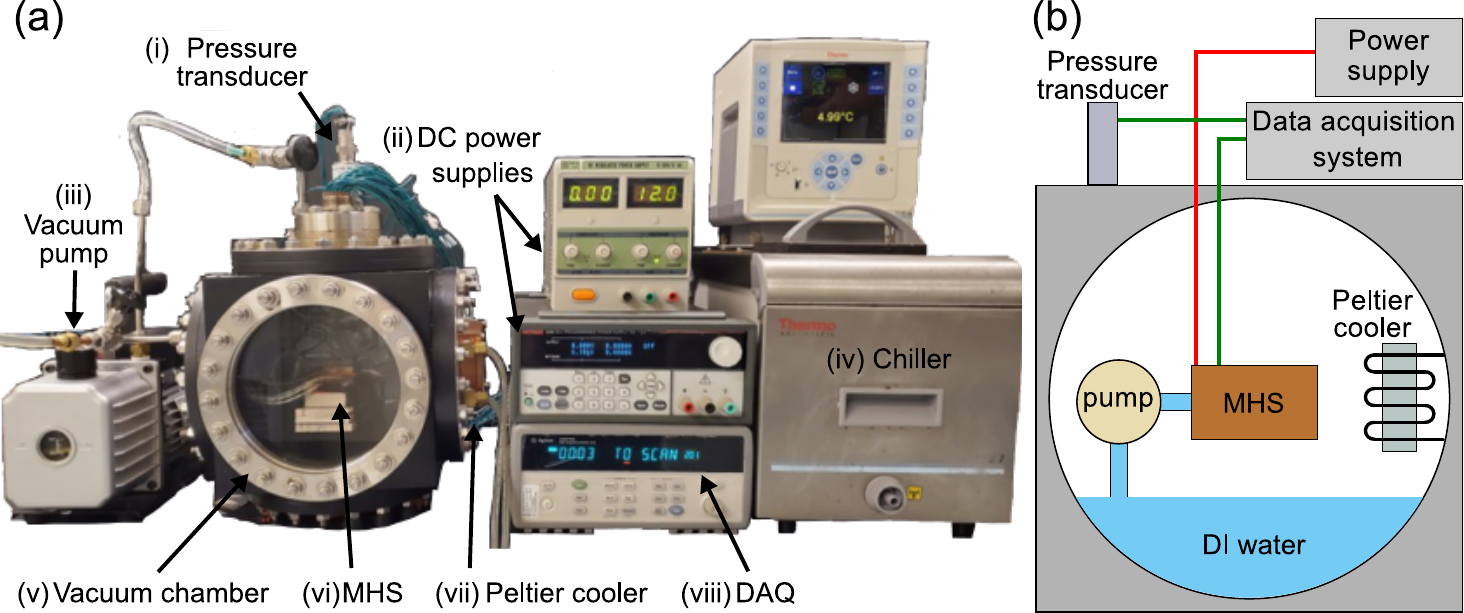}
	\caption{(a) Photograph of the experimental setup showing components of the experimental setup including (i) pressure transducer, (ii) DC power supplies, (iii) vacuum pump, (iv) chiller, (v) vacuum chamber, (vi) membrane assisted heat sink (MHS), (vii) peltier cooler, and (viii) data acquisition system (DAQ). (b) Schematic of the test loop.\label{fig:fig4}}
\end{figure*}
\subsection{Experimental setup and test section}
A photograph and schematic of the experimental setup is illustrated in \cref{fig:fig4}. The heat sink described in \cref{sec:sec3-1} is installed in a chamber made from a 6$\times$6$\times$6 in$^3$ (152.4$\times$152.4$\times$152.4 mm$^3$) stainless steel cube whose internal walls are coated with polytetrafluoroethylene (PTFE). Glass viewports are installed on two opposite flanges to allow for visualization of the apparatus while an electrical feedthrough on the top flange enables power delivery and sensor measurements whilst maintaining saturation conditions inside the chamber (16 kPa for the present experiments). Deionized (DI) water is used as the working liquid and is delivered to the heat sink by a piezoelectric micropump (Model MP6, manufactured by Bartels Mikrotechnik GmbH). Two pressure transducers are utilized to monitor the test chamber (Setra 730) and heat sink (Omega PX26) pressures. A thermoelectric cooler (TEC) installed on the chamber sidewall is used to condense vapor exiting the heat sink and a data acquisition system (DAQ) (Agilent Technologies 34970A) is used to record temperature and pressure data. During an experiment, the pressure inside the chamber ($P_{vapor}$) is reduced to 16 kPa using a vacuum pump. Next, heat is supplied to the test surface through the cartridge heaters and the working liquid inside the MHS is maintained at a constant absolute pressure $P_{pool}$. The induced device pressure in the MHS relative to the chamber is thus $\Delta{P} = P_{pool} - P_{vapor}$. The applied heat flux is increased incrementally in steps of 10 W/cm$^2$ until a sudden jump in the temperature is observed. The heat flux at this point is recorded to be the CHF at the specific device pressure P.
\subsection{Uncertainty analysis}
To determine the uncertainty in heat flux and heat transfer coefficient, the error and uncertainty associated with the thermocouples is considered. The uncertainty in determining the heat flux is caused by uncertainty in temperature readings, thermal conductivity of copper, and spacing between the thermocouples. \Cref{eqn:hfuncertain} was used to calculate the uncertainty in heat flux

\begin{equation} \label{eqn:hfuncertain}
\dfrac{\delta q''}{q''} = \sqrt{\left(\dfrac{\delta k}{k}\right)^{2} + \left(\dfrac{\delta \Delta T}{\Delta T}\right)^{2} + \left(\dfrac{\delta \Delta x}{\Delta x}\right)^{2}}
\end{equation}

where $\delta$k, $\delta\Delta$T, $\delta$x are the uncertainties in thermal conductivity, temperature gradient, and distance between thermocouples, respectively. Since $\Delta{T}=3T_3-4T_2+T_1$, $\delta\Delta$T can be calculated using \cref{eqn:delT}

\begin{equation} \label{eqn:delT}
\delta \Delta T = \sqrt{\left[ (3\delta T)^2 +(4\delta T)^2 +(\delta T)^2 \right]} \sim 5.09 \delta T
\end{equation}

The uncertainty associated with different experimental variables are tabulated in \cref{tab:uncertain}. The uncertainty in heat flux measurement is found to be $\pm$11.2\% at a lower value of 70 W/cm$^2$ and $\pm$7.1\% at the highest value of 700 W/cm$^2$. The uncertainty associated with calculating the heat transfer coefficient (HTC = $q/(A\times{T}_{sup})$), where \textit{q}, $T_{sup}$, and \textit{A} are the heat rate, surface superheat, and heater surface area, respectively, is estimated using \cref{eqn:HTC}

\begin{equation} \label{eqn:HTC}
\dfrac{\delta HTC}{HTC} = \sqrt{\left(\dfrac{\delta q}{q}\right)^{2} + \left(\dfrac{\delta T_{sup}}{T_{sup}}\right)^{2} + \left(\dfrac{\delta A}{A}\right)^{2}}
\end{equation}

where $\delta{q}$, $\delta{T}_{sup}$, and $\delta{A}$ are the errors in measuring the heat rate, surface superheat, and heater area, respectively. The uncertainty in HTC is determined to be $\pm$6.9\% at the lowest value of 310 kW/m2K and $\pm$33.5\% at the highest value of 890 kW/m2K.
\begin{table}[ht!]
\begin{center}
\caption{Uncertainties in measured experimental parameters \label{tab:uncertain}}
\begin{tabular}{c c}
\hline
Variable & Uncertainty\\
\hline
$\delta{T}/T$ & $\pm$0.65 K \\
\rowcolor{table-gray}
$\delta{k}/k$ & $\pm$2\% \\
$\delta\Delta{x}/\Delta{x}$ & $\pm$0.3\% \\
\rowcolor{table-gray}
$\delta{A}/A$ & $\pm$0.2\% \\
$\delta\Delta{P}/\Delta{P}$ & $\pm$0.25\% \\
\hline
\end{tabular}
\end{center}
\end{table}

\section{Results and Discussion \label{sec:results}}
\subsection{Critical heat flux in membrane based heat sinks}
\begin{figure}[ht!]
	\centering
	\includegraphics[width=0.48\textwidth]{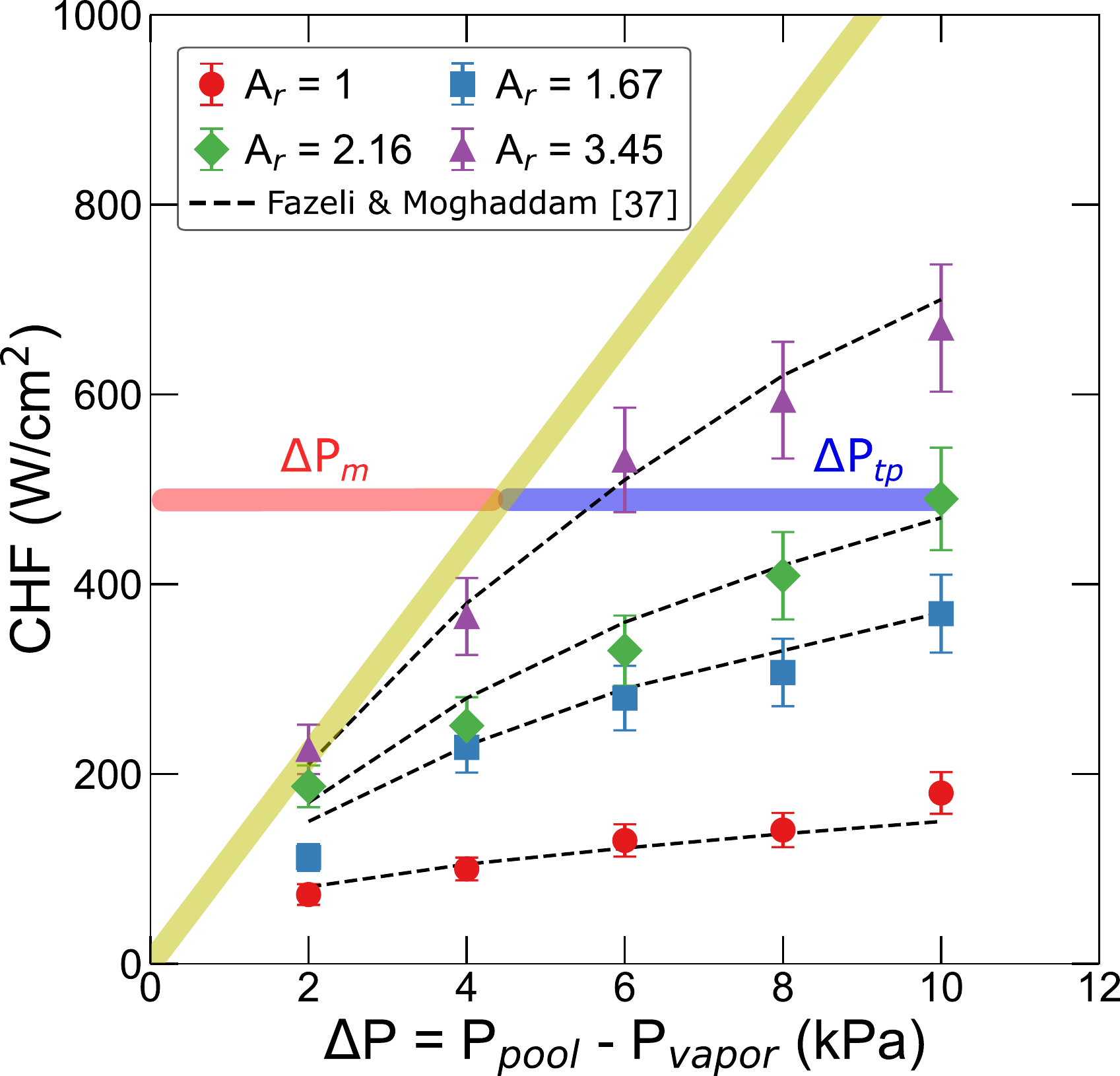}
	\caption{Critical heat flux (CHF) obtained on test surfaces with different $A_r$ as a function of device pressure $\Delta{P} = P_{pool} - P_{vapor}$. The dotted line represents the theoretical model described by Fazeli \& Moghaddam \cite{Fazeli2017} and fits the experimental data well. The yellow band represents the maximum heat flux based on the membrane permeability limit. The red and blue bands represent the contribution of membrane ($P_m$) and two-phase pressure drop ($P_{tp}$) to total device pressure drop ($\Delta{P}$), respectively. \label{fig:fig5}}
\end{figure}

The MHS performance was analyzed through a set of boiling tests conducted on all four surfaces listed in \cref{tab:tab1}, at device pressures ($\Delta{P} = P_{pool} - P_{vapor}$, see \cref{fig:fig2}) ranging from 2-10 kPa. To ensure repeatability, each test was repeated 3 times and the CHF reported in \cref{fig:fig5} is an average of the three tests. As described in \cref{sec:sec2}, an increase in device pressure $\Delta{P}$ reduces bubble residence time and vapor contact area on the heater surface, leading to enhanced liquid replenishment to the surface. Hence, we expect to see an increase in CHF with an increase in $\Delta{P}$. Further, an increase in $A_r$ is known to increase CHF \cite{Fazeli2017, Alipanah2020}. As evidenced in \cref{fig:fig5}, a marked increase in CHF is observed with an increase in surface area as well as device pressure, consistent with our prior studies \cite{Alipanah2020}. The yellow band in \cref{fig:fig5} denotes the maximum heat flux limit for a given device pressure based on membrane permeability. Consistent with results obtained at atmospheric pressure \cite{Fazeli2017, Alipanah2020, Tamvada2021}, above a certain pressure ($\sim$2 kPa), CHF is not limited by the membrane permeability, but by surface properties. It can also be inferred from \cref{fig:fig5} that the rate of CHF enhancement with increase in $\Delta{P}$ declines at higher pressures.
The dashed lines in Fig. 5 represents a semi-empirical correlation developed by Fazeli \& Moghaddam \cite{Fazeli2017}. The model accounts for effect of wickability ($\mathfrak{W}$), effective heat transfer area ($A_{r,\epsilon}$), and liquid pressure on CHF. Test results at atmospheric pressure \cite{Fazeli2017} and here suggest that on structures with similar wickability, increasing surface area almost linearly enhances the CHF limit; hence, it is expected that $q"_{CHF} \propto A_r$. It has also been demonstrated that structures with higher wickability can facilitate higher CHF values at low $\Delta{P}$, and that the impact of wickability and $A_r$ is directly dependent on $\Delta{P}$. Accordingly, \cref{eqn:chfmodel} represents the additive and multiplicative effect of wickability and heat transfer area, respectively, on CHF:
\begin{multline} \label{eqn:chfmodel}
q''_{CHF}(A_{r,\epsilon}, \Delta{P}, \mathfrak{W}) = \left[ q''_{nw}(A_{r,\epsilon}=1, \Delta{P}\sim 0, \mathfrak{W}\sim 0) \right. \\  \left. + q''_w(A_{r,\epsilon}=1, \Delta{P}, \mathfrak{W}) \right] f\left(A_{r,\epsilon}, \Delta{P}\right)
\end{multline}
where $A_{r,\epsilon}$ denotes effective heat transfer area which is determined by considering the thermal effectiveness of surface structures (viz., $A_r$, with its thermal efficiency accounted for), and $\mathfrak{W}$ denotes surface wickability. Here, $q"_{nw}$ represents CHF on a plain surface (i.e. $A_{r,\epsilon}=1, \mathfrak{W}\sim 0$) at $\Delta{P}\sim 0$. Therefore, its value is independent of the surface wickability and heat transfer area and can only be changed by altering the liquid contact angle. $q"_w$, on the other hand, represents heat flux associated with the wicking process and changes with $\mathfrak{W}$ and $\Delta{P}$. Finally, \textit{f} denotes the effect of enhanced heat transfer area as well as liquid pressure on surface structure effectiveness. \Cref{fig:fig5} shows a close agreement of the experimental data with the model, suggesting that when CHF is not limited by membrane permeability, effective heat transfer area and wickability contribute additively and multiplicatively as described by Fazeli \& Moghaddam \cite{Fazeli2017} even at lower pressures. It is important to note that since the model only factors in the effect of surface structures and fluid properties, it is solely valid below the membrane transport limit.
\subsection{Heat transfer coefficient (HTC)}
\begin{figure}[ht!]
	\centering
	\includegraphics[width=0.48\textwidth]{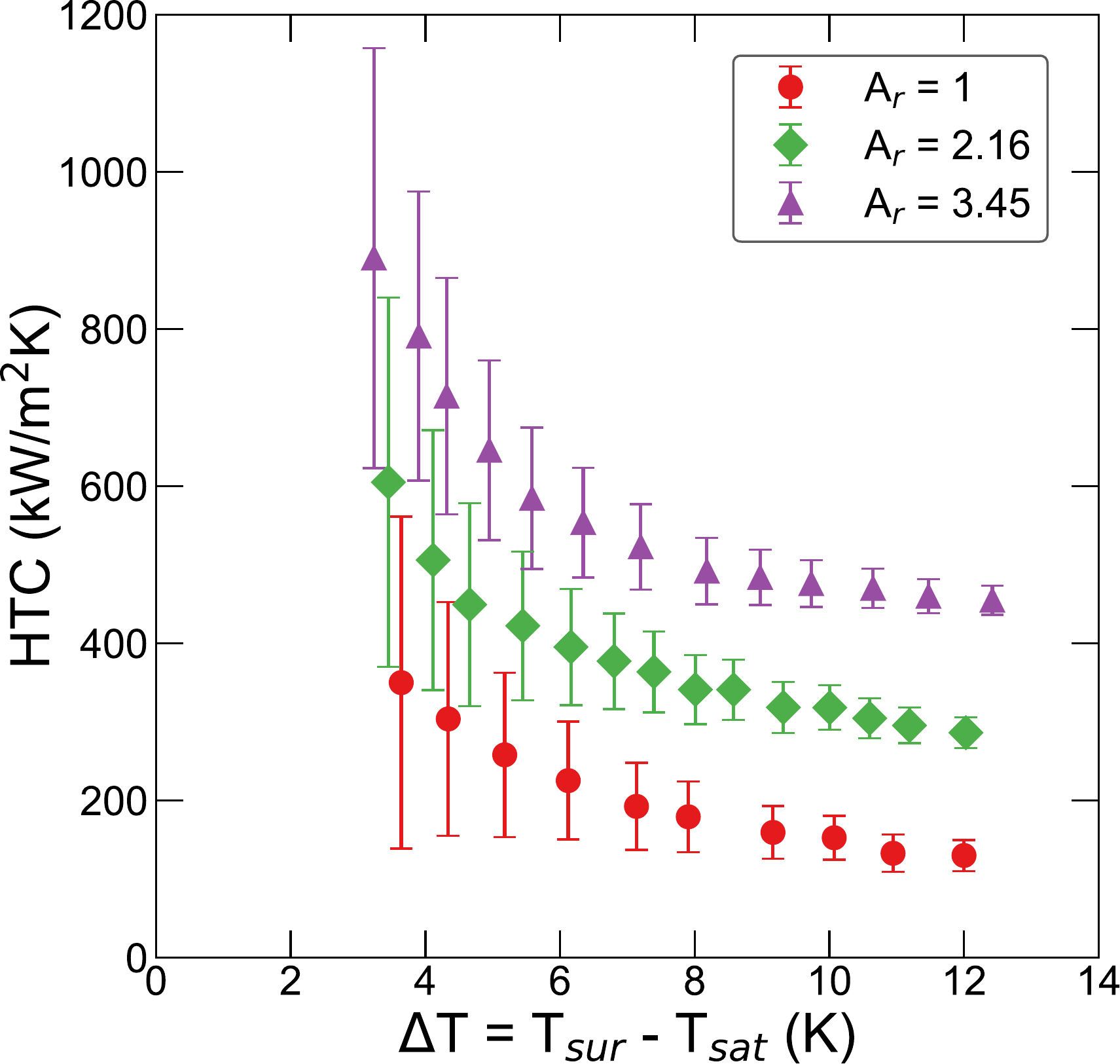}
	\caption{Heat transfer coefficient (HTC) as a function of surface superheat for surfaces with different enhanced area ratios ($A_r$ = 1, 2.16, 3.45) operated at a pressure potential of 8 kPa.\label{fig:fig6}}
\end{figure}

\begin{figure}[htp!]
	\centering
	\includegraphics[width=0.48\textwidth]{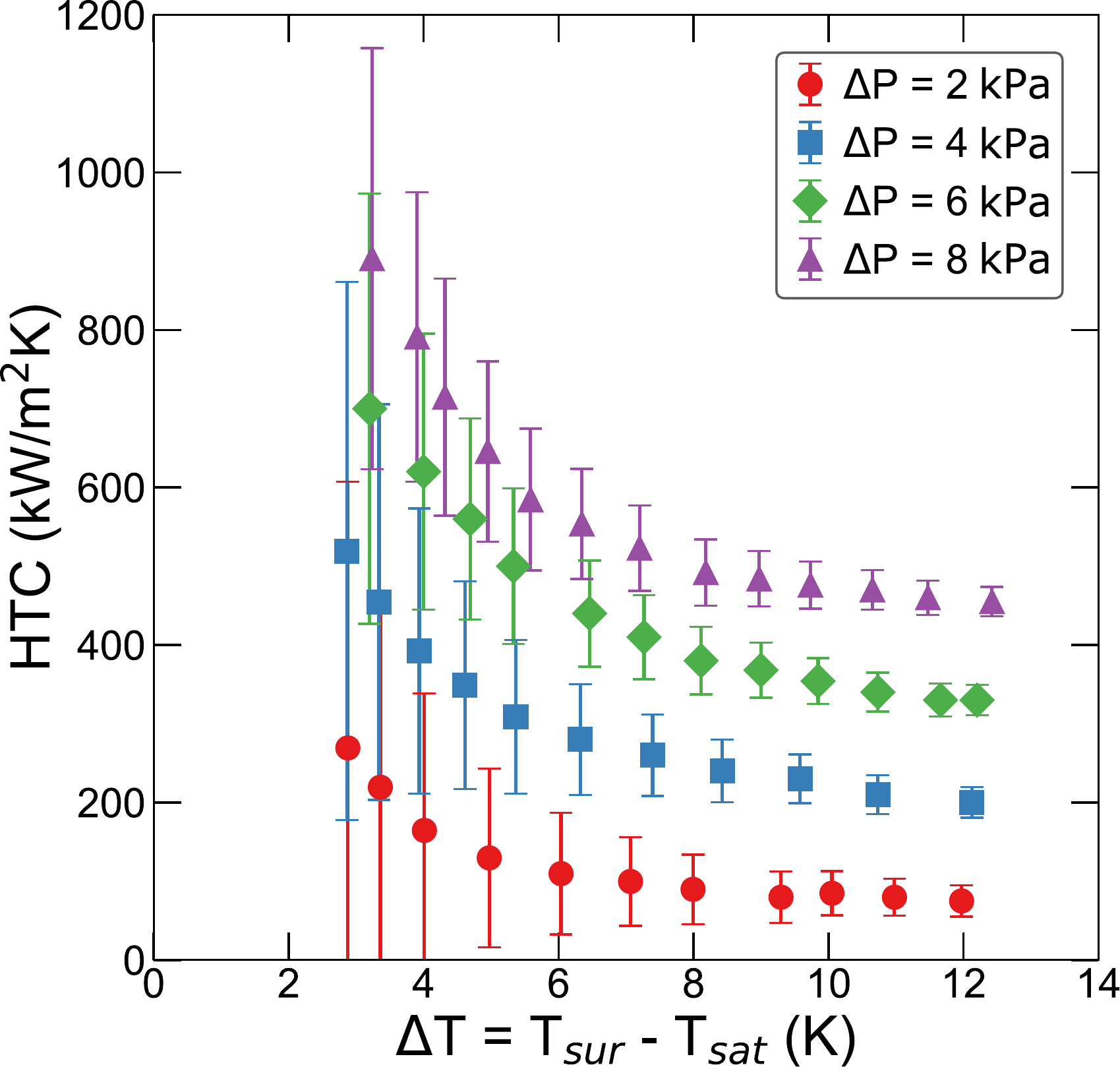}
	\caption{Heat transfer coefficient (HTC) as a function of surface superheat for various device pressures $\Delta{P} = P_{pool} – P{vapor}$ on a surface with $A_r$ = 3.45.\label{fig:fig7}}
\end{figure}

As discussed in \cref{sec:intro}, cooling in DCs demand efficient heat transfer as it is directly tied to their cost of operation. We thus evaluate the heat transfer coefficient HTC = $q\/(A\times\Delta{T})$ of MHS, where q is the heat rate, A is the heat transfer area, and $\Delta{T}$ (= $T_{sur} – T_{sat}$) is the surface superheat. \Cref{fig:fig6} plots HTC as a function of $\Delta{T}$ for tests conducted at $\Delta{P}$ = 8 kPa on surfaces with different $A_r$. For a given surface, HTC is found to decrease with an increase in $\Delta{T}$, eventually reaching a near-constant value beyond a superheat of 10$^\circ$C. This suggests that an initial increase in heat flux occurs at the expense of heat transfer efficiency; however, at later stages, no significant detriment in HTC is observed. Furthermore, a trend of increase in HTC with increase in $A_r$ is observed for a given $\Delta{T}$, corroborating previous reports that surface structures enable higher heat transfer coefficients through increased heat transfer area \cite{Fazeli2017}.
Next, to investigate the effect of device pressure on HTC, we consider a surface with $A_r$ = 3.45. HTC is plotted as a function of $\Delta{T}$ in \cref{fig:fig7} and it is observed that as $\Delta{P}$ increases, HTC increases for a given $\Delta{T}$. This is consistent with the MHS operating principle, since an increase in device pressure improves surface rewetting leading to maximal heat transfer through phase change. Surface structures and induced pressure thus increase HTC as envisaged, and a maximum heat transfer coefficient of 890 kW/m$^2$K is obtained on a surface with $A_r$ = 3.45 at a $\Delta{P}$ of 8 kPa.
Although an increase in heat flux reduces HTC at higher superheats, an efficient vapor generation and discharge process facilitated by the membrane allows even high heat fluxes of 580.4 W/cm$^2$ to be dissipated at an HTC of 502.6 kW/m$^2$K. 

\begin{figure}[ht!]
	\centering
	\includegraphics[width=0.48\textwidth]{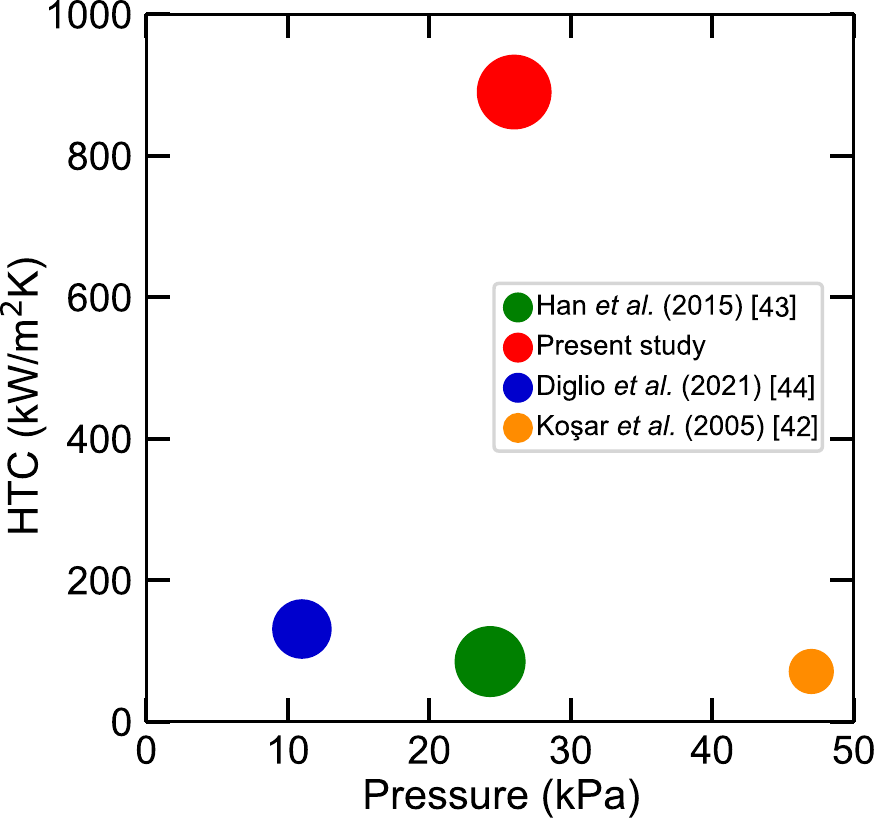}
	\caption{Comparison of heat transfer coefficient (HTC) obtained by MHS with other heat sinks operated using water at lower pressures \cite{Kosar2005, Han2016, Diglio2021}. The pressure indicated here is the liquid saturation pressure. The size of the bubbles represents the heat flux at which the heat transfer coefficient is obtained.\label{fig:fig8}}
\end{figure}

The maximum HTC obtained using MHS is compared with existing two-phase cooling technologies operated using water at comparable system pressures and plotted in \cref{fig:fig8} \cite{Kosar2005, Han2016, Diglio2021}. The pressure reported in \cref{fig:fig8} is the liquid saturation pressure where the HTC is reported while the size of the bubbles represents the heat flux (W/cm$^2$) at which the HTC is obtained. We find that reports of phase-change heat sinks operated using water at sub-ambient pressures is relatively sparse. Han et al. \cite{Han2016}, who utilized pin-fin structures and reported a maximum HTC of 85 kW/m$^2$K at a liquid saturation pressure of 28 kPa, while Diglio \textit{et al.} \cite{Diglio2021} recently demonstrated that HTCs of 131 kW/m$^2$K can be dissipated through direct device impingement (DDI). Kosar \textit{et al.} \cite{Kosar2005} studied boiling in a microchannel with reentrant cavities and reported a maximum HTC of 71 kW/m$^2$K. In comparison, MHS achieves a maximum HTC of 890 kW/m$^2$K (at a heat flux of 324 W/cm$^2$), roughly an order of magnitude higher than other two-phase heat sinks studied at comparable pressures with water as the working fluid. We also note that the performance of MHS drastically decreases when operated at sub-ambient system pressures (cf. Refs. \cite{Fazeli2017, Alipanah2020}).
\subsection{Surface temperature}
\begin{figure}[ht!]
	\centering
	\includegraphics[width=0.48\textwidth]{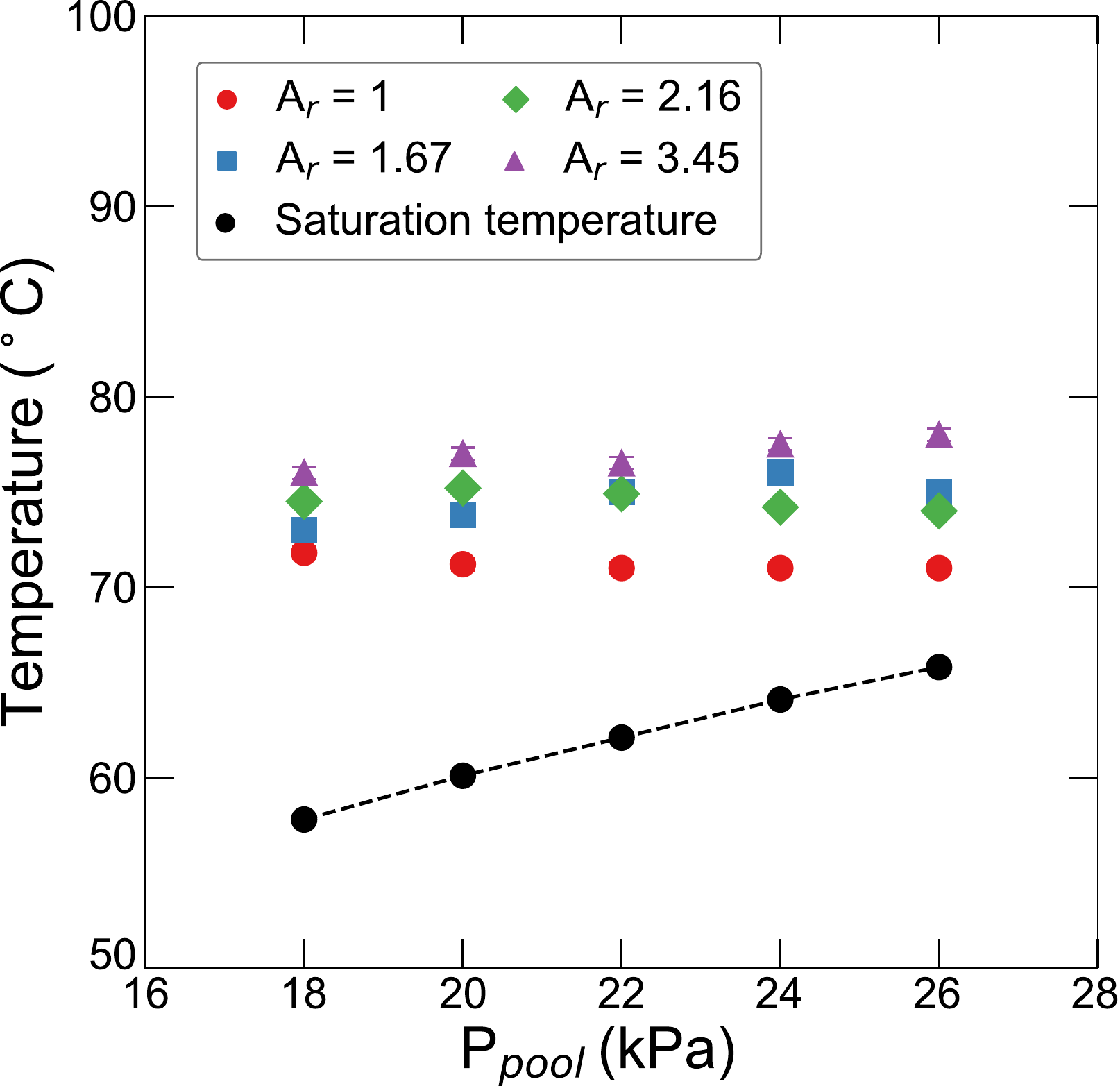}
	\caption{Surface temperatures of test surfaces with different Ar as a function of liquid pressure ($P_{pool}$), plotted alongside saturation temperature of water (dashed line). Surface temperatures do not vary appreciably with varying device pressure which results in reduced superheat at higher pressures.\label{fig:fig9}}
\end{figure}

Due to constantly varying heat loads in a DC, it is desirable for a heat sink to dissipate a range of fluxes without significant temperature variations i.e., maintaining stable surface temperatures below silicon junction temperature is of paramount importance to ensure safe and reliable operation of electronics. The experimental data points in \cref{fig:fig9} correspond to the distinct test cases presented in \cref{fig:fig5} for all test surfaces and device pressures. The saturation temperature of water for the corresponding $P_{pool}$ is also plotted for comparison. In conventional heat sinks, increasing the heat flux often results in a higher pressure drop and change in the saturation temperature of the fluid, and consequently the heater surface temperature. In contrast, membrane assisted vapor removal augmented by induced liquid pressure transpires as a decrease in surface superheat with an increase in device pressure.

\subsection{Analysis of pressure distribution within the heat sink}
Pressure drop is an important performance characteristic of MHS that has not been analyzed in our prior publications. To understand the underlying physics of pressure drop, we have considered various factors contributing to the pressure drop when CHF is not limited by the membrane permeability. Pressure drop in MHS is caused by 3 factors - (i) liquid pressure drop at the inlet channel and partly over the heater surface, (ii) pressure drop due to phase change ($\Delta{P}_{tp}$), and (iii) vapor pressure drop through the membrane ($\Delta{P}_{m}$). Liquid enters the heat sink through a channel of cross section 1$\times$1 mm$^2$ and expands into an enclosure with lateral cross-sectional area of 1$\times$7 mm$^2$ (see \cref{fig:fig3}(b)). This pressure drop in the channel, estimated using the Darcy-Weisbach equation, is negligible ($\sim$100 Pa) \cite{Bergman2011}. Therefore, the major contributors towards pressure drop are $\Delta{P}_{tp}$ and $\Delta{P}_{m}$. The individual contribution of these factors can be inferred from \cref{fig:fig5}. For example, considering the test on a surface with $A_r$ = 2.16 at a $\Delta{P}$ of 10 kPa, $\Delta{P}_{m}$ is represented by the red band while $\Delta{P}_{tp}$ is represented by the blue band.
We suppose that $\Delta{P}_{tp}$ is due to the effect of growing vapor bubbles during boiling that is commonly known to push against the incoming fluid to the heat sink \cite{Mikic1970}. Hence, an increase in two-phase pressure drop is expected with increasing nucleation. This is evident when considering the pressure drop on the plain surface ($A_r$ = 1) – the contribution of two-phase pressure drop increases with an increase in heat flux. Evidently, $\Delta{P}_{tp}$ is the highest on this surface even though its CHF is much less than those of the structured surfaces. We believe that increase in frequency and density of stable vapor columns on structured surfaces is responsible for this unique pressure drop characteristic of the MHS. Surface structures greatly enhance direct evaporation of liquid films, feeding into vapor columns. In comparison with nucleation, the pressure drop across these stable vapor columns is significantly lower since these columns do not rapidly expand. However, an increase in heat flux through direct evaporation (i.e., thin film and contact line evaporation) leads to a higher vapor generation rate resulting in a higher pressure drop through the membrane. These trends are all evident in \cref{fig:fig5}; as the surface structures area is increased, the contribution of two-phase pressure drop ($\Delta{P}_{tp}$) decreases while the contribution of membrane pressure drop ($\Delta{P}_{m}$) increases.

\section{Implications for future data centers}
\subsection{Replacing existing technologies}
\begin{figure}[ht!]
	\centering
	\includegraphics[width=0.48\textwidth]{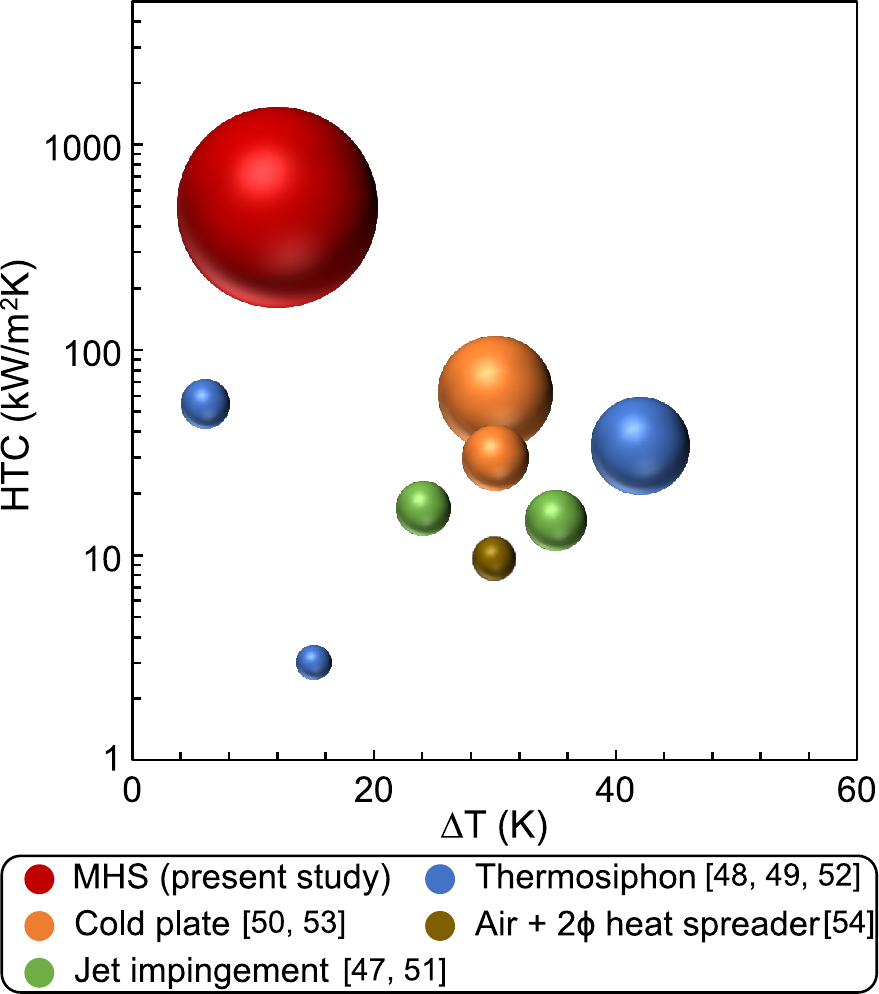}
	\caption{Comparison of different cooling strategies for electronics. Cold plates \cite{Hoang2020, Coldplate}, thermosiphons \cite{Lamaison2017, Panse2017, Amalfi2020}, jet impingement strategies \cite{Qiu2015, Kwon2020}, and air cooled heat sinks with 2-phase heat spreaders (vapor chambers) \cite{Vapor} are compared with membrane assisted heat sinks (MHS). The heat transfer coefficient (HTC) is plotted as a function of the surface superheat ($\Delta$T) while the size of the bubbles represents the heat flux at which the HTC is obtained. Membrane assisted heat sinks (MHS) exhibit an HTC at least an order of magnitude higher than air cooled heat sinks \cite{Qiu2015, Lamaison2017, Panse2017, Hoang2020, Kwon2020, Amalfi2020, Coldplate, Vapor}.\label{fig:fig10}}
\end{figure}

\Cref{fig:fig10} shows the state-of-the-art performance of various passive and active technologies including thermosiphons, cold plates, jet impingement, etc. \cite{Qiu2015, Lamaison2017, Panse2017, Hoang2020, Kwon2020, Amalfi2020, Coldplate, Vapor} in comparison with MHS, in the form of the HTC plotted against surface superheat, with the size of the bubbles representing the heat flux at which the HTC is obtained.
It is important to note that comparing these vastly different technologies using a common benchmark is complex and does not convey a holistic picture about the technologies since they are constructed for different application needs. Some applications demand removal of high heat fluxes while others necessitate low surface temperatures, and some others require a low operating power. However, the HTC (accounting for $\Delta{T}$) is a crucial metric to DC cooling due to its direct impact on DC operating expenditure (OpEx) and can be evaluated to determine favorable technologies to be employed. We choose technologies that are closest in their working conditions to DC environment and exclude system level parameters such as pumping power, pressure drop, etc. in our analysis. It is clear that MHS outperforms other technologies by exhibiting a HTC of 500 kW/m$^2$K at a heat flux of 580 W/cm$^2$. Although current power densities do not largely exceed this level \cite{Rack}, MHS promises a future-readiness without the need for frequent capital expenditure (CapEx) in the form of constant replacement of cooling systems at the rack level.
\subsection{System level integration benefits}
\begin{figure*}[h!]
	\centering
	\includegraphics[width=0.75\textwidth]{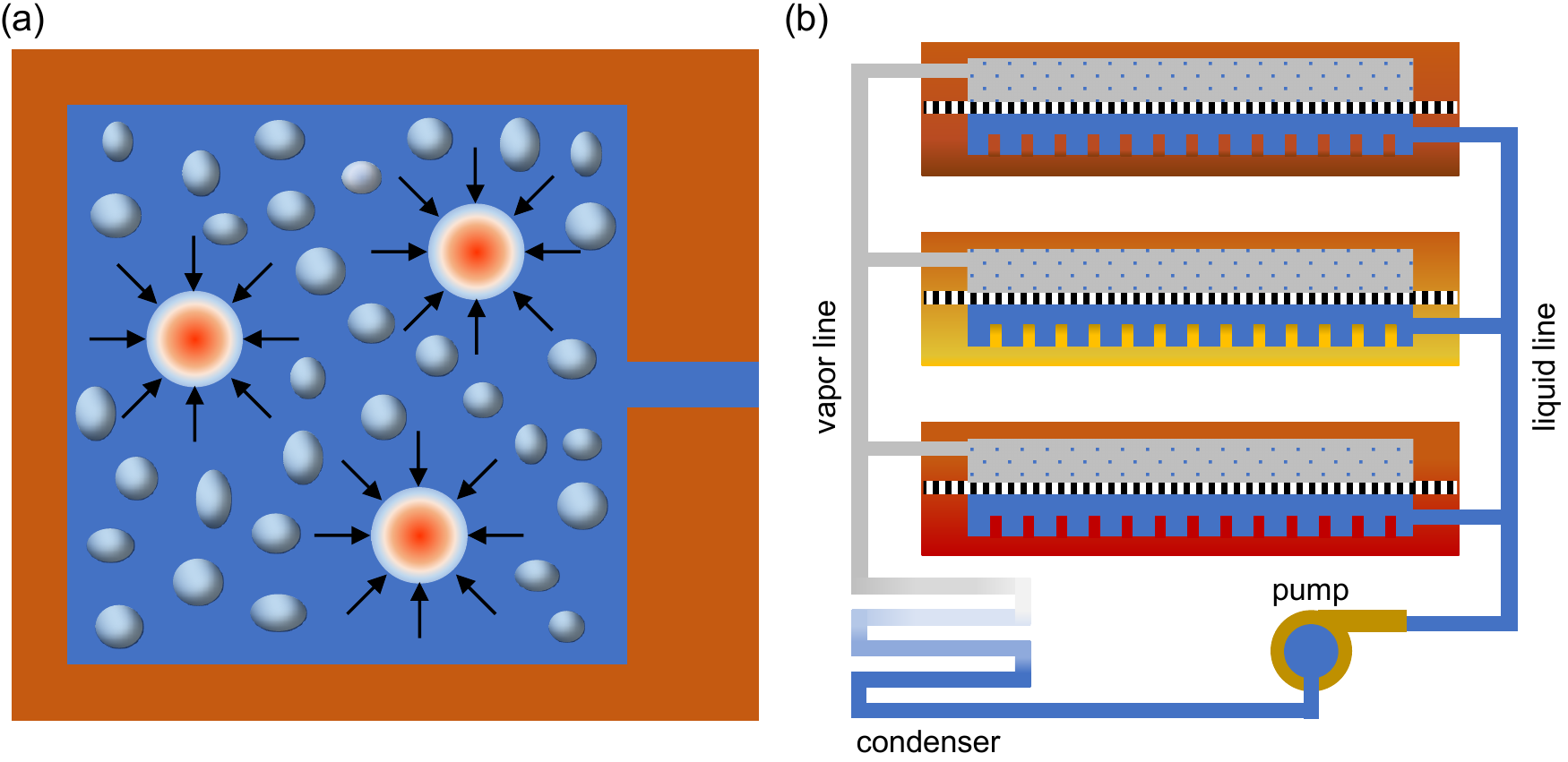}
	\caption{A schematic showing the benefits of system level integration of MHS. (a) Any hotspots in the heat sinks will be quenched as an imposed pressure promotes liquid supply to the hotspots. (b) Prospect of multiple heat sinks at different heat loads being connected on a common liquid supply and vapor exit line. \label{fig:fig11}}
\end{figure*}
Further, from the analysis presented in \cref{sec:results}, the system level benefits of incorporating MHS into a DC become apparent. \Cref{fig:fig11} shows a schematic representation of the system level benefits associated with MHS. Due to the architecture of MHS which imposes a positive omnidirectional pressure on the liquid (see \cref{fig:fig2}(b)), any hotspots generated on the chip surface is quickly mitigated through the rushing of the liquid to the hotspots (see \cref{fig:fig11}(a)) as opposed to the common problem of vapor backflow suffered by conventional microchannel heat sinks \cite{Hedau2022}. Secondly, MHSs exhibit a self-regulating pressure control due to the presence of the membrane, which can be leveraged to connect multiple heat sinks through a common liquid and vapor loop (see \cref{fig:fig11}(b)). Consider three heat sinks under different heat loads connected through a common liquid and vapor line as shown in \cref{fig:fig11}(b).

Liquid will automatically be diverted to the heat sink with the most heat load without the need for mass controllers whilst maintaining a 100\% vapor quality. Heat sinks at lower heat loads would also be maintained at 100\% vapor quality due to the constant liquid pressure in the liquid line and the membrane, enabling a self-regulating mechanism. The vapor line can be used to collect and condense vapor from all heat sinks to be directed back to the heat sinks.

\section{Conclusions}
This work provides a comprehensive analysis of the membrane assisted heat sink performance characteristics at operating conditions pertinent to DCs. The results show a substantial decline in CHF at low operating pressures relative to operation at the atmospheric pressure. Our experiments reveal that CHF and HTC increase with an increase in device pressure (Ppool) and enhanced area ratio ($A_r$), and we report a maximum CHF and HTC of 670 W/cm2 and close to $\sim$ 1 MW/m$^2$K, respectively. We predicted CHF using the model proposed by Fazeli and Moghaddam, considering the effect of wickability, effective heat transfer area, and liquid pressure. MHS exhibits great surface temperature controllability over a wide range of operating conditions and unique pressure drop characteristics. Results presented here not only imply an exceptional proposition for the use of MHS in high performance DCs, but also raise performance benchmarks for phase change heat sinks. Lastly, we expound system level benefits of integrating MHS in DCs such as hotspot mitigation and self-regulating loop control.

\section{Acknowledgements}
This work was supported by the National Science Foundation grant number \href{https://www.nsf.gov/awardsearch/showAward?AWD_ID=1934354}{CBET - 1934354} with Dr. Ying Sun as program manager. The authors thank Dr. Morteza Alipanah and Mr. Sourabh Kakade for assistance with experiments.

\bibliography{export.bib}

\begin{thebibliography}{10}
\expandafter\ifx\csname url\endcsname\relax
  \def\url#1{\texttt{#1}}\fi
\expandafter\ifx\csname urlprefix\endcsname\relax\def\urlprefix{URL }\fi
\expandafter\ifx\csname href\endcsname\relax
  \def\href#1#2{#2} \def\path#1{#1}\fi

\bibitem{Li2020}
J.~Li, J.~Jurasz, H.~Li, W.~Q. Tao, Y.~Duan, J.~Yan, A new indicator for a fair
  comparison on the energy performance of data centers, Applied Energy 276
  (2020) 115497.
\newblock \href {https://doi.org/10.1016/J.APENERGY.2020.115497}
  {\path{doi:10.1016/J.APENERGY.2020.115497}}.

\bibitem{Kheirabadi2016}
A.~C. Kheirabadi, D.~Groulx, Cooling of server electronics: A design review of
  existing technology, Applied Thermal Engineering 105 (2016) 622--638.
\newblock \href {https://doi.org/10.1016/J.APPLTHERMALENG.2016.03.056}
  {\path{doi:10.1016/J.APPLTHERMALENG.2016.03.056}}.

\bibitem{Alkharabsheh2015}
S.~Alkharabsheh, J.~Fernandes, B.~Gebrehiwot, D.~Agonafer, K.~Ghose, A.~Ortega,
  Y.~Joshi, B.~Sammakia, A brief overview of recent developments in thermal
  management in data centers, Journal of Electronic Packaging 137 (12 2015).
\newblock \href {https://doi.org/10.1115/1.4031326}
  {\path{doi:10.1115/1.4031326}}.

\bibitem{Garimella2013}
S.~V. Garimella, T.~Persoons, J.~Weibel, L.-T. Yeh, Technological drivers in
  data centers and telecom systems: Multiscale thermal, electrical, and energy
  management, Applied Energy 107 (2013) 66--80.
\newblock \href {https://doi.org/10.1016/j.apenergy.2013.02.047}
  {\path{doi:10.1016/j.apenergy.2013.02.047}}.

\bibitem{Dayarathna2016}
M.~Dayarathna, Y.~Wen, R.~Fan, Data center energy consumption modeling{:} a
  survey, IEEE Communications Surveys \& Tutorials 18 (2016) 732--794.
\newblock \href {https://doi.org/10.1109/COMST.2015.2481183}
  {\path{doi:10.1109/COMST.2015.2481183}}.

\bibitem{Chu2019}
W.~X. Chu, C.~C. Wang, A review on airflow management in data centers, Applied
  Energy 240 (2019) 84--119.
\newblock \href {https://doi.org/10.1016/J.APENERGY.2019.02.041}
  {\path{doi:10.1016/J.APENERGY.2019.02.041}}.

\bibitem{Marcinichen2014}
J.~B. Marcinichen, D.~Wu, S.~Paredes, J.~R. Thome, B.~Michel, Dynamic flow
  control and performance comparison of different concepts of two-phase on-chip
  cooling cycles, Applied Energy 114 (2014) 179--191.
\newblock \href {https://doi.org/10.1016/J.APENERGY.2013.09.018}
  {\path{doi:10.1016/J.APENERGY.2013.09.018}}.

\bibitem{Kandasamy2022}
R.~Kandasamy, J.~Y. Ho, P.~Liu, T.~N. Wong, K.~C. Toh, S.~J. Chua, Two-phase
  spray cooling for high ambient temperature data centers: Evaluation of system
  performance, Applied Energy 305 (2022) 117816.
\newblock \href {https://doi.org/10.1016/J.APENERGY.2021.117816}
  {\path{doi:10.1016/J.APENERGY.2021.117816}}.

\bibitem{Cui2017}
Y.~Cui, C.~Ingalz, T.~Gao, A.~Heydari, Total cost of ownership model for data
  center technology evaluation, Proceedings of the 16th InterSociety Conference
  on Thermal and Thermomechanical Phenomena in Electronic Systems, ITherm 2017
  (2017) 936--942\href {https://doi.org/10.1109/ITHERM.2017.7992587}
  {\path{doi:10.1109/ITHERM.2017.7992587}}.

\bibitem{Rasmussen2011}
N.~Rasmussen,
  \href{http://52.2.195.45/images/6_cost_of_ownership_p3.pdf}{Determining total
  cost of ownership for data center and network room infrastructure},
  Relat{\'o}rio t{\'e}cnico, Schneider Electric, Paris 8 (2011).
\newline\urlprefix\url{http://52.2.195.45/images/6_cost_of_ownership_p3.pdf}

\bibitem{Matin2020}
M.~H. Matin, S.~Moghaddam, Thin liquid films formation and evaporation
  mechanisms around elongated bubbles in rectangular cross-section
  microchannels, International Journal of Heat and Mass Transfer 163 (2020)
  120474.
\newblock \href {https://doi.org/10.1016/j.ijheatmasstransfer.2020.120474}
  {\path{doi:10.1016/j.ijheatmasstransfer.2020.120474}}.

\bibitem{Moghaddam2006}
S.~Moghaddam, \href{http://drum.lib.umd.edu/handle/1903/9139}{Microscale study
  of nucleation process in boiling of low-surface-tension liquids} (2006).
\newline\urlprefix\url{http://drum.lib.umd.edu/handle/1903/9139}

\bibitem{Moghaddam2009}
S.~Moghaddam, K.~Kiger, Physical mechanisms of heat transfer during single
  bubble nucleate boiling of fc-72 under saturation conditions-i. experimental
  investigation, International Journal of Heat and Mass Transfer 52 (2009)
  1284--1294.
\newblock \href {https://doi.org/10.1016/j.ijheatmasstransfer.2008.08.018}
  {\path{doi:10.1016/j.ijheatmasstransfer.2008.08.018}}.

\bibitem{Matin2021}
M.~H. Matin, S.~Moghaddam, Mechanism of transition from elongated bubbles to
  wavy-annular regime in flow boiling through microchannels, International
  Journal of Heat and Mass Transfer 176 (2021) 121464.
\newblock \href {https://doi.org/10.1016/j.ijheatmasstransfer.2021.121464}
  {\path{doi:10.1016/j.ijheatmasstransfer.2021.121464}}.

\bibitem{Verma2021}
A.~Verma, N.~Kumar, R.~Raj, Direct prediction of foamability of aqueous
  surfactant solutions using property values, Journal of Molecular Liquids 323
  (2021) 114635.
\newblock \href {https://doi.org/10.1016/J.MOLLIQ.2020.114635}
  {\path{doi:10.1016/J.MOLLIQ.2020.114635}}.

\bibitem{Xi2004}
Y.~Xi, E.~F. Schubert, Junction-temperature measurement in gan ultraviolet
  light-emitting diodes using diode forward voltage method, Applied Physics
  Letters 85 (2004) 2163--2165.
\newblock \href {https://doi.org/10.1063/1.1795351}
  {\path{doi:10.1063/1.1795351}}.

\bibitem{Zajaczkowski2016}
Experimental verification of heat transfer coefficient for nucleate boiling at
  sub-atmospheric pressure and small heat fluxes, Heat and Mass
  Transfer/Waerme- und Stoffuebertragung 52 (2016) 205--215.
\newblock \href {https://doi.org/10.1007/s00231-015-1549-8}
  {\path{doi:10.1007/s00231-015-1549-8}}.

\bibitem{Michaie2017}
S.~Michaie, R.~Rulli{\`e}re, J.~Bonjour, Experimental study of bubble dynamics
  of isolated bubbles in water pool boiling at subatmospheric pressures,
  Experimental Thermal and Fluid Science 87 (2017) 117--128.
\newblock \href {https://doi.org/10.1016/j.expthermflusci.2017.04.030}
  {\path{doi:10.1016/j.expthermflusci.2017.04.030}}.

\bibitem{McGillis1991}
W.~R. McGillis, V.~P. Carey, J.~S. Fitch, W.~R. Hamburgen, Pool boiling
  enhancement techniques for water at low pressure, Proceedings - IEEE
  Semiconductor Thermal and Temperature Measurement Symposium (1991)
  64--72\href {https://doi.org/10.1109/STHERM.1991.152914}
  {\path{doi:10.1109/STHERM.1991.152914}}.

\bibitem{Nukiyama1934}
S.~Nukiyama, The maximum and minimum values of the heat q transmitted from
  metal to boiling water under atmospheric pressure, Journal of the Society of
  Mechanical Engineers 37 (1934) 367--374.
\newblock \href {https://doi.org/10.1299/JSMEMAGAZINE.37.206_367}
  {\path{doi:10.1299/JSMEMAGAZINE.37.206_367}}.

\bibitem{Bonilla1941}
C.~Bonilla, Heat transmission to boiling binary liquid mixtures, Transactions
  of AIChE 37 (1941) 685--705.

\bibitem{Cichelli1945}
M.~T. Cichelli, C.~F. Bonilla, Heat transfer to liquids boiling under pressure,
  Transactions of the American institute of chemical engineers 41~(6) (1945)
  755--787.

\bibitem{Kutateladze1948}
S.~S. Kutateladze,
  \href{https://access.library.oregonstate.edu/pdf/1035856.pdf}{On the
  transition to film boiling under natural convection}, Kotloturbostroenie 3
  (1948) 20.
\newline\urlprefix\url{https://access.library.oregonstate.edu/pdf/1035856.pdf}

\bibitem{Rohsenow1956}
W.~M. P.~Rohsenow, Griffith, Correlation of maximum heat transfer data for
  boiling of saturated liquids, Chem. Eng. Prog. (1956) 47.

\bibitem{Zuber1959}
N.~Zuber, Hydrodynamic aspects of boiling heat transfer (thesis) (6 1959).
\newblock \href {https://doi.org/10.2172/4175511} {\path{doi:10.2172/4175511}}.

\bibitem{Moissis1963}
R.~Moissis, P.~J. Berenson, On the hydrodynamic transitions in nucleate
  boiling, Journal of Heat Transfer 85 (1963) 221--226.
\newblock \href {https://doi.org/10.1115/1.3686075}
  {\path{doi:10.1115/1.3686075}}.

\bibitem{Bui1985}
T.~D. Bui, V.~K. Dhir, Transition boiling heat transfer on a vertical surface,
  Journal of Heat Transfer 107 (1985) 756--763.
\newblock \href {https://doi.org/10.1115/1.3247501}
  {\path{doi:10.1115/1.3247501}}.

\bibitem{Kandlikar2001}
S.~G. Kandlikar, A theoretical model to predict pool boiling chf incorporating
  effects of contact angle and orientation, Journal of Heat Transfer 123 (2001)
  1071--1079.
\newblock \href {https://doi.org/10.1115/1.1409265}
  {\path{doi:10.1115/1.1409265}}.

\bibitem{Attinger2014}
D.~Attinger, C.~Frankiewicz, A.~R. Betz, T.~M. Schutzius, R.~Ganguly, A.~Das,
  C.-J. Kim, C.~M. Megaridis, Surface engineering for phase change heat
  transfer: A review, MRS Energy \& Sustainability 1 (2014) 1--40.
\newblock \href {https://doi.org/10.1557/mre.2014.9}
  {\path{doi:10.1557/mre.2014.9}}.

\bibitem{Jo2011}
H.~Jo, H.~S. Ahn, S.~Kang, M.~H. Kim, A study of nucleate boiling heat transfer
  on hydrophilic, hydrophobic and heterogeneous wetting surfaces, International
  Journal of Heat and Mass Transfer 54 (2011) 5643--5652.
\newblock \href {https://doi.org/10.1016/j.ijheatmasstransfer.2011.06.001}
  {\path{doi:10.1016/j.ijheatmasstransfer.2011.06.001}}.

\bibitem{Rahman2014}
M.~M. Rahman, E.~{\"O}l{\c c}ero{\u g}lu, M.~McCarthy, Role of wickability on
  the critical heat flux of structured superhydrophilic surfaces, Langmuir 30
  (2014) 11225--11234.
\newblock \href {https://doi.org/10.1021/la5030923}
  {\path{doi:10.1021/la5030923}}.

\bibitem{Chu2011}
K.~H. Chu, R.~Enright, E.~N. Wang, Microstructured surfaces for enhanced pool
  boiling heat transfer, Vol.~10, 2011, pp. 679--685.
\newblock \href {https://doi.org/10.1115/imece2011-65169}
  {\path{doi:10.1115/imece2011-65169}}.

\bibitem{Chu2013}
K.~H. Chu, Y.~S. Joung, R.~Enright, C.~R. Buie, E.~N. Wang, Hierarchically
  structured surfaces for boiling critical heat flux enhancement, Applied
  Physics Letters 102 (2013).
\newblock \href {https://doi.org/10.1063/1.4801811}
  {\path{doi:10.1063/1.4801811}}.

\bibitem{Kandlikar2017}
S.~G. Kandlikar, Enhanced macroconvection mechanism with separate liquid-vapor
  pathways to improve pool boiling performance, Journal of Heat Transfer 139 (5
  2017).
\newblock \href {https://doi.org/10.1115/1.4035247/384465}
  {\path{doi:10.1115/1.4035247/384465}}.

\bibitem{Moghaddam2019}
S.~Moghaddam, S.~A. Fazeli,
  \href{https://patents.google.com/patent/US10492333B2/en}{U.s. patent no.
  10,492,333} (11 2019).
\newline\urlprefix\url{https://patents.google.com/patent/US10492333B2/en}

\bibitem{Moghaddam2021}
S.~Moghaddam, S.~A. Fazeli,
  \href{https://patents.google.com/patent/US10897833B2/en}{U.s. patent no.
  10,897,833} (1 2021).
\newline\urlprefix\url{https://patents.google.com/patent/US10897833B2/en}

\bibitem{Fazeli2017}
A.~Fazeli, S.~Moghaddam, A new paradigm for understanding and enhancing the
  critical heat flux (chf) limit, Scientific Reports 7 (2017) 5184.
\newblock \href {https://doi.org/10.1038/s41598-017-05036-2}
  {\path{doi:10.1038/s41598-017-05036-2}}.

\bibitem{Alipanah2020}
M.~Alipanah, S.~Moghaddam, Ultra-low pressure drop membrane-based heat sink
  with 1000 w/cm$^2$ cooling capacity and 100\% exit vapor quality,
  International Journal of Heat and Mass Transfer 161 (2020) 120312.
\newblock \href {https://doi.org/10.1016/j.ijheatmasstransfer.2020.120312}
  {\path{doi:10.1016/j.ijheatmasstransfer.2020.120312}}.

\bibitem{Kutaleladse1951}
S.~Kutaleladse, A hydrodynamic theory of changes in the boiling process under
  free convection conditions, Izv. Akad. Nauk., USSR, Otd. Tekh. Nauk 4 (1951)
  529--529.

\bibitem{Tamvada2021}
S.~R. Tamvada, M.~Alipanah, S.~Moghaddam, Membrane-based two phase heat sinks
  for high heat flux electronics and lasers, IEEE Transactions on Components,
  Packaging and Manufacturing Technology (2021) 1--1\href
  {https://doi.org/10.1109/TCPMT.2021.3115419}
  {\path{doi:10.1109/TCPMT.2021.3115419}}.

\bibitem{Zhou2003}
G.~Zhou, J.~C. Yang, Temperature effect on the cu 2 o oxide morphology created
  by oxidation of cu(0 0 1) as investigated by in situ uhv tem, Applied Surface
  Science 210 (2003) 165--170.
\newblock \href {https://doi.org/10.1016/S0169-4332(03)00159-4}
  {\path{doi:10.1016/S0169-4332(03)00159-4}}.

\bibitem{Kosar2005}
Reduced pressure boiling heat transfer in rectangular microchannels with
  interconnected reentrant cavities.

\bibitem{Han2016}
X.~Han, Y.~Joshi, A.~Fedorov, Flow boiling of water at sub-atmospheric pressure
  in staggered micro pin-fin heat sink, Proceedings of the 15th InterSociety
  Conference on Thermal and Thermomechanical Phenomena in Electronic Systems,
  ITherm 2016 (2016) 799--804\href
  {https://doi.org/10.1109/ITHERM.2016.7517628}
  {\path{doi:10.1109/ITHERM.2016.7517628}}.

\bibitem{Diglio2021}
P.~Diglio, D.~Shia, P.~Tadayon, D.~Kulkarni, J.-Y. Chang, Performances of
  two-phase cooling technologies that uses water as working fluid under
  sub-ambient pressures, IEEE, 2021, pp. 86--92.
\newblock \href {https://doi.org/10.1109/ITherm51669.2021.9503281}
  {\path{doi:10.1109/ITherm51669.2021.9503281}}.

\bibitem{Bergman2011}
T.~Bergman, F.~Incropera, D.~DeWitt, A.~Lavine, Fundamentals of heat and mass
  transfer, 2011.

\bibitem{Mikic1970}
B.~B. Mikic, W.~M. Rohsenow, P.~Griffith, On bubble growth rates, International
  Journal of Heat and Mass Transfer 13 (1970) 657--666.
\newblock \href {https://doi.org/10.1016/0017-9310(70)90040-2}
  {\path{doi:10.1016/0017-9310(70)90040-2}}.

\bibitem{Qiu2015}
L.~Qiu, S.~Dubey, F.~H. Choo, F.~Duan, Recent developments of jet impingement
  nucleate boiling, International Journal of Heat and Mass Transfer 89 (2015)
  42--58.
\newblock \href {https://doi.org/10.1016/J.IJHEATMASSTRANSFER.2015.05.025}
  {\path{doi:10.1016/J.IJHEATMASSTRANSFER.2015.05.025}}.

\bibitem{Lamaison2017}
N.~Lamaison, C.~L. Ong, J.~B. Marcinichen, J.~R. Thome, Two-phase
  mini-thermosyphon electronics cooling: Dynamic modeling, experimental
  validation and application to 2u servers, Applied Thermal Engineering 110
  (2017) 481--494.
\newblock \href {https://doi.org/10.1016/j.applthermaleng.2016.08.198}
  {\path{doi:10.1016/j.applthermaleng.2016.08.198}}.

\bibitem{Panse2017}
S.~S. Panse, S.~G. Kandlikar, A thermosiphon loop for high heat flux removal
  using flow boiling of ethanol in omm with taper, International Journal of
  Heat and Mass Transfer 106 (2017) 546--557.
\newblock \href {https://doi.org/10.1016/J.IJHEATMASSTRANSFER.2016.09.020}
  {\path{doi:10.1016/J.IJHEATMASSTRANSFER.2016.09.020}}.

\bibitem{Hoang2020}
C.~H. Hoang, S.~Khalili, B.~Ramakrisnan, S.~Rangarajan, Y.~Hadad, V.~Radmard,
  K.~Sikka, S.~Schiffres, B.~Sammakia, An experimental apparatus for two-phase
  cooling of high heat flux application using an impinging cold plate and
  dielectric coolant, 36th Annual Semiconductor Thermal Measurement, Modeling
  and Management Symposium, SEMI-THERM 2020 - Proceedings (2020) 32--38\href
  {https://doi.org/10.23919/SEMI-THERM50369.2020.9142831}
  {\path{doi:10.23919/SEMI-THERM50369.2020.9142831}}.

\bibitem{Kwon2020}
B.~Kwon, T.~Foulkes, T.~Yang, N.~Miljkovic, W.~P. King, Air jet impingement
  cooling of electronic devices using additively manufactured nozzles, IEEE
  Transactions on Components, Packaging and Manufacturing Technology 10 (2020)
  220--229.
\newblock \href {https://doi.org/10.1109/TCPMT.2019.2936852}
  {\path{doi:10.1109/TCPMT.2019.2936852}}.

\bibitem{Amalfi2020}
R.~L. Amalfi, F.~Cataldo, J.~B. Marcinichen, J.~R. Thome, Experimental
  characterization of a server-level thermosyphon for high-heat flux
  dissipations, Vol. 2020-July, IEEE Computer Society, 2020, pp. 402--409.
\newblock \href {https://doi.org/10.1109/ITherm45881.2020.9190186}
  {\path{doi:10.1109/ITherm45881.2020.9190186}}.

\bibitem{Coldplate}
B.~Corporation.
\newblock
  \href{https://www.boydcorp.com/thermal/liquid-cooling/liquid-cold-plate.html}{Liquid
  cold plates overview} [online] (2022).

\bibitem{Vapor}
B.~Corporation,
  \href{https://www.boydcorp.com/thermal/two-phase-cooling/vapor-chambers/3d-vapor-chamber-assemblies.html}{3d
  vapor chamber assemblies} (2022).
\newline\urlprefix\url{https://www.boydcorp.com/thermal/two-phase-cooling/vapor-chambers/3d-vapor-chamber-assemblies.html}

\bibitem{Rack}
U.~institute Blog,
  \href{https://journal.uptimeinstitute.com/rack-density-is-rising/}{Rack
  density is rising} (2022).
\newline\urlprefix\url{https://journal.uptimeinstitute.com/rack-density-is-rising/}

\bibitem{Hedau2022}
G.~Hedau, R.~Raj, S.~K. Saha, Complete suppression of flow boing instability in
  microchannel heat sinks using a combination of inlet restrictor and flexible
  dampener, International Journal of Heat and Mass Transfer 182 (2022) 121937.
\newblock \href {https://doi.org/10.1016/J.IJHEATMASSTRANSFER.2021.121937}
  {\path{doi:10.1016/J.IJHEATMASSTRANSFER.2021.121937}}.

\bibitem{Van1976}
S.~Van~Stralen, W.~Sluyter, R.~Cole, Bubble growth rates in nucleate boiling of
  aqueous binary systems at subatmospheric pressures, International Journal of
  Heat and Mass Transfer 19~(8) (1976) 931--941.
\newblock \href {https://doi.org/10.1016/0017-9310(76)90205-2}
  {\path{doi:10.1016/0017-9310(76)90205-2}}.

\end{thebibliography}
\bibliographystyle{elsarticle-num}
\nocite{*}
\end{document}